\documentclass[structabstract]{aa}
\usepackage[utf8]{inputenc}
\usepackage[T1]{fontenc}
\usepackage{lmodern}
\usepackage{graphicx}
\usepackage{verbatim}
\usepackage{subfigure}
\usepackage{textcomp}
\usepackage{epsfig}
\usepackage{natbib}
\usepackage{amssymb,amsmath}

\bibpunct{(}{)}{;}{a}{}{,}

\begin{document}

\title{Effect of angular opening on the dynamics of relativistic hydro jets}
\author{Remi Monceau-Baroux \inst{1} \and Rony Keppens \inst{1} \and Zakaria Meliani \inst{2}}
\institute{Centre for mathematical Plasma Astrophysics, Department of Mathematics, KU Leuven, \hfill \\
Celestijnenlaan 200B, 3001 Heverlee, Belgium \and LUTh, Observatoire de Paris}


\abstract{Relativistic jets emerging from active galactic nuclei (AGN) cores transfer energy from the core of the AGN to their surrounding interstellar/intergalactic medium through shock-related and hydrodynamic instability mechanisms. Because jets are observed to have finite opening angles, one needs to quantify the role of conical versus cylindrical jet propagation in this energy transfer.}{We adopt parameters representative for Faranoff-Riley class II AGN jets with finite opening angles. We study how such an opening angle affects the overall dynamics of the jet and its interaction with its surrounding medium and therefore how it influences the energy transfer between the AGN and the external medium. We also point out how the characteristics of this external medium, such as its density profile, play a role in the dynamics.}{This study exploits our parallel adaptive mesh refinement code MPI-AMRVAC with its special relativistic hydrodynamic model, incorporating an equation of state with varying effective polytropic index. We initially studied mildly underdense jets up to opening angles of 10 degrees, at Lorentz factors of about 10, inspired by input parameters derived from observations. Instantaneous quantifications of the various interstellar medium (ISM) volumes affected by jet injection and their energy content allows one to quantify the role of mixing versus shock-heated cocoon regions over the simulated time intervals.}{We show that a wider opening angle jet results in a faster deceleration of the jet and leads to a wider radial expansion zone dominated by Kelvin-Helmholtz and Rayleigh-Taylor instabilities. The energy transfer mainly occurs in the shocked ISM region by both the frontal bow shock and cocoon-traversing shock waves, in a roughly 3 to 1 ratio to the energy transfer of the mixing zone, for a 5 degree opening angle jet. The formation of knots along the jet may be related to X-ray emission blobs known from observations. A rarefaction wave induces a dynamically formed layered structure of the jet beam.}{Finite opening angle jets can efficiently transfer significant fractions (25 \% up to 70 \%) of their injected energy over a growing region of shocked ISM matter. The role of the ISM stratification is prominent for determining the overall volume that is affected by relativistic jet injection. While our current 2D simulations give us clear insights into the propagation characteristics of finite opening angle, hydrodynamic relativistic jets, we need to expand this work to 3D.}
\maketitle

\section{Introduction}

Relativistic plasma jets have been exhaustively studied and observed in detail in radio galaxies such as Abell 2052, Hercules A, or Cygnus A \citep{Krause2005, Krawczynski2012}. From multi-wavelength observations in specific sources we have understood important aspects of the ways in which jets are emitted by active galactic nuclei (AGNs) and form pronounced radio lobes \citep{blandford1974,scheuer1973,cohen2007,Lister2009}.

While observations give access to dynamic parameters such as the Lorentz factor $\gamma$ or an estimate of the energy flux from the overall luminosity, uncertainties remain in the properties of the jet matter itself, for instance, its composition, density or pressure and the precise topology and strength of the magnetic field. Simulation is a powerful tool to make estimates these parameters, and one can study different scenarios with them, e.g. the composition of the jet consisting of mainly electrons and protons, or electrons and positrons \citep{kundt1980}.

Many astrophysical relativistic jets have a high Lorentz factor, with $\gamma$ varying from 5 to 30 for AGN jets \citep{kellermann2004}. Therefore, relativistic dynamics models are necessary to study them.

Jets associated with AGNs have been separated into different classes. Among the observational classification of radioloud AGN, we find weaksource Fanaroff-Riley class I objects (FR-I, \cite{fanaroff1974}), with a brighter luminosity near the origin of the jet. An example is the jet in the 3C 296 object \citep{laing2006}. Powerful sources are known as Faranoff-Riley class II radio galaxies (FR-II) and lobe-dominated radio-loud quasars. An example is the jet in the 3C216 object \citep{fejes1992}. In all cases, the jet has been shown to have a relativistic velocity with a Lorentz factor of up to 5 for FR-I and a Lorentz factor of several dozens for FR-II \citep{M.Marti1997,cohen2007,Lister2009}. We concentrate here on FR-II jet parameters.

Previous works have shown how a jet can be launched on opening magnetic field configurations \citep{blandford1982}. For relativistic magnetized winds from a central object, additional collimation by disk winds may be needed for their ultimate collimation, see \citet{bogovalov2005}. Those works illustrate how hard it is to obtain a fully collimated jet with an overall cylindrical geometry. Observations clearly show finite opening angles of jets, e.g. \citet{puskarev2011,zezas2005,frazer2000}. Flaring models that fit observations well were made for FR-I jets, as in the work of \citet{laing2006}. In this model, the original angle obtained from collimation in the source increases due to internal characteristics. For FR-II jets, although the beam still shows a finite width and opening angle, this conical geometry is often discarded by taking a wider width of the beam, or assuming a precession of the source \citep{S.M.O'Neill2005}. Few models invoke an actual conical geometry up to a few degrees \citep{Z.Meliani2008}. The cylindrical injection models dominate because FR-II jets are believed to be quickly re-collimated by internal shocks. Because we studied jet propagation beyond 0.5 parsec away from the central object, we ignored general relativistic as well as magnetic effects and performed pure hydro studies for varying opening angles. We assumed axisymmetry of the jet and therefore used 2D simulations. This allowed us to explore parametrically hydrodynamic relativistic simulations. Nevertheless, a more realistic study would imply 3D simulations, which we will perform in forthcoming studies.

The interest in studing relativistic jets associated with AGNs has increased recently, since jets feed back energy into the interstellar/intergalactic (ISM/IGM) medium, which makes them candidate feedback processes at play in the larger scale formation of galaxy clusters. For that reason we need to quantify the energy transfer efficiencies of AGN jets for cylindrical and conical jets.

Multiple mechanisms are thought to be responsible for the energy transfer between the jet and its surroundings. Direct energy conversion takes place at the end of the jet beam, immediately beyond the Mach disk surface of the jet (in a hydro case), where kinetic energy becomes thermal energy through efficient relativistic shocks. Hydrodynamic instabilities taking place at the interface between the shocked jet and shocked ISM matter mix the two fluids, allowing for energy transfer from the now hot outer part of the beam to the surroundings of the jet. Additionally, a bow shock forms around the entire jet cocoon, propagating in all directions and transferring energy to further regions of space. We quantify the relevance of each and how a finite opening angle at the jet inlet affects these energy transfers. 

Many other dynamical features are known from observations, for instance the existence of X-ray blobs along the jet path. Different scenarios are studied to explain this emission, such as particle acceleration at bow shocks when stars cross the beam of the jet. We propose that our jet dynamics model to explain those blobs without invoking external agents such as star-related bow shocks. This explanation revisits the role of reconfinement shocks to explain these knots, as in the work by \cite{komissarov1997}.

In the second part of this article we describe the relativistic hydrodynamic model and the way we use it with our code, MPI-AMRVAC. We discuss our settings and initial conditions and their motivation. Part 3 will compare our different simulations by giving a parametric study of the effect of an opening angle on the dynamics of our jets. This comparison studies the geometry and energy content of the jets. The fourth part discusses two aspects found in our simulations that relate to the internal structure of the jet:  a scenario for X-ray blobs and the dynamically formed, layered structure of the jet. We finish with general conclusions obtained from this work.

\section{Simulation}
\subsection{Governing equations}
\label{equation}

Like in the previous work of \citet{Z.Meliani2008}, we assumed axisymmetry. We use  a relativistic hydrodynamic model one with the relativistic variant of the Euler equations with the Mathews approximation \citep{mathews1976} to the Synge gas equation as a closure.
They express the conservation of the different relevant quantities. The first equation is the conservation of particle number 
\begin{center}
\begin{equation}
\frac{\partial \rho \gamma}{\partial t} + \vec{\nabla} \cdot (\rho \gamma \vec{v}) = 0\,,
\label{eulerdensity}
\end{equation} 
\end{center}
where $\gamma$ is the Lorentz factor, $\rho$ is the proper density and $\vec{v}$ is the three-velocity of the fluid in the chosen lab frame. 
The second equation is the conservation of momentum, written as
\begin{center}
\begin{equation}
\frac{\partial \vec{S}}{\partial t} + \vec{\nabla} \cdot (\vec{S} \vec{v}) + \vec{\nabla} p = 0 \,,
\label{eulermomentum}
\end{equation} 
\end{center}
where $p$ is the pressure in the fluid frame and $\vec{S}$ is the momentum density defined as
\begin{center}
\begin{equation}
\vec{S} = h \gamma^2 \rho \vec{v} \,.
\end{equation}
\end{center}
The relativistic specific enthalpy $h$ is defined as
\begin{center}
\begin{equation}
h = \frac{1}{2}\left((\Gamma+1)\frac{e}{m_p}-(\Gamma-1)\frac{m_p}{e}\right) ,
\end{equation} 
\end{center}
where $\Gamma$ is a parameter, taken as the non-relativistic value of the polytropic index of the gas $\frac{5}{3}$, $e = m_p +e_{th}$ is the specific internal energy including the rest (proton) mass $m_p$ and the specific thermal energy $e_{th}$.
This is equivalent to using a spatially varying, effective $\Gamma_{\mathrm{eff}} =\Gamma - \frac{\Gamma-1}{2}(1 - \frac{m_p^2}{e^2})$ that varies between its classical value of ${5}/{3}$ and its relativistic value of ${4}/{3}$.
The third equation expresses the conservation of energy density
\begin{center}
\begin{equation}
\frac{\partial \tau}{\partial t} + \vec{\nabla} \cdot (\vec{S} - \rho \gamma \vec{v}) = 0 \,,
\label{eulerenergy}
\end{equation} 
\end{center}
where the energy density $\tau$ in the lab frame with the rest mass contribution already subtracted, is written as
\begin{center}
\begin{equation}
\tau = \rho h \gamma^2 - p - \rho \gamma\,.
\end{equation}
\end{center}
The Mathews approximation to the Synge gas equation is our closure relation, and it writes as
\begin{center}
\begin{equation}
p = \left(\frac{\Gamma - 1}{2}\right) \rho \left(\frac{e}{m_p} - \frac{m_p}{e}\right)\,.
\label{syngeequation}
\end{equation}
\end{center} 
The speed of light is taken equal to unity in our model and therefore in all equations above.

\subsection{Code and implementation}

The code we used is MPI-AMRVAC\footnote{Code homepage at \textit{\textquoteleft homes.esat.kuleuven.be/\textasciitilde keppens'}} \citep{vac2012}, designed to solve a set of equations in conservative form, as is the case for our relativistic HD model. 
The solver is a hybrid version of the HLLC scheme, for which a validation can be found in the work of \citet{Z.Meliani2008}. The novelty of this scheme is that it switches to a TVDLF solver where the HLLC solver could induce spurious oscillations.
More validation tests for relativistic HD and MHD can be found on the code website and in the work of \citet{vac2012}.

The code incorporates block-tree adaptive mesh refinement (AMR), and we used a weighted combination of the normalized second derivative of the density variable $\rho\gamma$, the pressure $p$ and the Lorentz factor to follow the dynamics of the jet and automatically refine the grid where dynamics takes place. Additionally, a forced refinement of the grid was activated around the inlet region of the jet, up to the maximum level specified, to accurately handle the bottom boundary.

\subsection{Initial conditions}
\label{IC}

The rectangular domain was defined to be 20 by 5 pc jet beam radius in length and width, respectively $(z,r)$, with a length normalization of 1 parsec which corresponds to 400 by 100 jet beam radius as we show below. The domain did not include the central engine that launches the jet, but started at $z_0=0.5$ distance from the source. therefore we can avoid treating the close surroundings of an AGN core, which requires general relativistic theory. From a numerical point of view, all AMR sets start with a base resolution of 300 by 190 grid points, in $(z,r)$ respectively, allowing four levels of refinement (hence an effective resolution of 2400 by 1520). We made sure that the resolution allowed us to resolve the jet properly: the jet width on inlet, fixed to be $r_b=0.05$ parsec, is 15 grid points wide at the maximal level of refinement taht is enforced in the jet region.
We set the density on the inlet of the jet $\rho_{\mathrm{jet}}$ such that the jet total kinetic luminosity is $L_{\mathrm{jet}}=10^{46}$ erg.s$^{-1}$ with a Lorentz factor $\gamma$ equal to 10, in a jet with radius $r_b$ given above, all values representative for an FR-II type AGN jet.

For the ISM we chose a model similar to the beta model of the King atmosphere \citep{king1962} profile. This assumes that the ISM has been structured into a density variation by the winds coming from the central object \citep{S.M.O'Neill2005,Krause2005}, resulting in a slowly decreasing density inside a core radius. Outside of this radius the density drops rapidly up to an asymptotic value. This ISM density then follows
\begin{equation}
\rho_{\mathrm{ISM}} = \rho_0\left(1+\frac{z^2}{z_\mathrm{c}^2}\right)^{-\beta}, 
\label{kingatmo}
\end{equation}
where $z_c$ is setting the size of the core of the galactic source, $\rho_0$ quantifies the density at the source of the jet ($\rho_0 = 5000$ g.$\mathrm{cm}^{-3}$ for cases A to E and $\rho_0=500$ g.$\mathrm{cm}^{-3}$ for cases F to H), and $\beta$ is a parameter set here to 0.75. The distance $z$ is here the axial direction instead of the actual radial distance from the central engine. This was done for numerical convenience, because it more easily handles the bottom boundary and avoids an artificial 'pixelization' of the density background at low refinement levels. We moreover assumed that the jet will not propagate over a large radial distance, resulting in a distance to the source $d_{source}\simeq z$ for $r$ small. Normally, $z_c$ is set to a few kpc, corresponding to the average size of a galaxy. This would result in a flat density profile in our parsec scale simulation. This is the case for simulations A and B, which were injected in uniform media, denoted \textquoteleft uniform' in Table \ref{parameter}. For simulations C, D and E we set $z_c$ to be 0.5 to investigate a possible change in the propagation of the jet as it becomes more dense than the ISM. With this setting in the domain the density profile behaves as a simple decreasing power law. This is denoted as \textquoteleft King' in the summary Table \ref{parameter}. Finally, simulations F, G and H are in between with a $z_c$ at 5 pc, called \textquoteleft King II' in the summary table.

\begin{table}[position]
\begin{center}
\begin{tabular}{|p{2.8cm}|p{2.8cm}|p{1.85cm}|}
\hline 
\textbf{Beam luminosity} & \textbf{Domain size (pc)} & \textbf{Domain resolution}   \\ 
\hline 
$10^{46} erg.s^{-1}$ & [0.5 20]x[0 5] & 2400x1520  \\ 
\hline 
\textbf{Beam radius (pc)} & \textbf{Beam $\gamma$} &    \\ 
\hline 
0.05 & 10 &    \\ 
\hline 
\end{tabular} 
\end{center}
   \caption{General parameters for all simulations.}
\end{table}

\begin{table}[position]
\begin{center}
\begin{tabular}{|p{1.0cm}|p{2.4cm}|p{1.3cm}|p{1.15cm}|}
\hline \textbf{Case} & \textbf{Density ratio} & \textbf{Opening angle} & \textbf{Density profile } \\ 
\hline 
\hline 
A & $0.01$ & 5\textdegree & uniform\\ 
\hline 
B & $0.0001$ & 5\textdegree & uniform\\ 
\hline 
C & $0.018$ & 0\textdegree & King\\ 
\hline 
D & $0.018$ & 5\textdegree &  King\\ 
\hline 
E & $0.018$ & 10\textdegree &  King\\ 
 \hline
F & $0.1$ & 0\textdegree &  King II\\ 
\hline 
G & $0.1$ & 5\textdegree &  King II\\ 
\hline 
H & $0.1$ & 10\textdegree &  King II\\ 
 \hline
\end{tabular} 
\end{center}
   \caption{Parameters distinguish between the simulations.}
   	\label{parameter}
\end{table}

The non-dimensional numeric value for the pressure was set on the whole domain to be uniformly small ($p_0 = 10^{-3} \rho_0$ for A to E and $p_0 = 10^{-2} \rho_0$ for cases F to H) so that it does not dominate the early stage of the dynamics. Having a uniform pressure also removes the need to have a gravity potential to keep the ISM in force balance.
We call $\eta$ the density ratio between the jet and the ISM, i.e. $\rho_{\mathrm{jet}}/\rho_{\mathrm{ISM}}=\eta$ at the position $z=z_0$. 

We chose for cases A and B a ratio $\eta$ equal to 0.01 and 0.0001, respectively, investigating the dynamics of an underdense jet. For cases C, D and E, this ratio was taken to be $\eta=0.018$. For those three cases the density of the ISM decreases rapidly in the domain and we expect the density of the jet and the ISM to match around the end of the domain. Cases F, G and H have a ratio  $\eta=0.1$ and a \textquoteleft core' inside the domain, resulting in our simulation box in a less rapidly decreasing density profile of the ISM, the density of both jet and ISM match around the middle of the domain.

\begin{figure}
\begin{center}
	\includegraphics[height=.3\textheight]{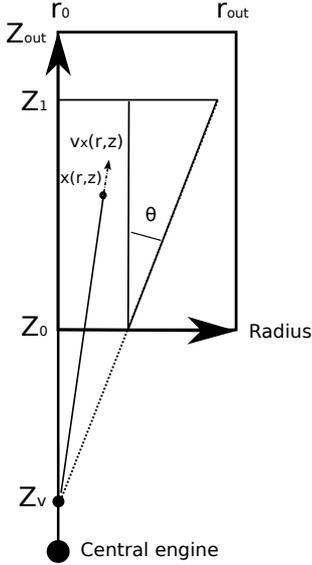}
	\caption{Geometry of the jet as used in the simulation: the jet enters the domain at $z_0=0.5$ pc from the source with a width setting $r_b$ as calculated assuming a first angle of 5\textdegree. A flaring-like case assumes a fixed angle $\theta$, which is added to a reference cylindrical jet, thereby introducing a `virtual source location'. The actual domain of the simulation is comprised between $z_0$ and $z_{out}$ for the axial direction and $r=0$ and $r_{out}$ for the radial direction. The velocity field is initiated in a finite region up to $z_1$, according to the formulae given in the text.}
	\label{jetgeo}
\end{center}
\end{figure}

We simulated on domains $z\in[z_0,z_{\mathrm{out}}]$ and $r\in[0,r_{\mathrm{out}}]$.
The boundaries at $z_{\mathrm{out}}$ and $r_{\mathrm{out}}$, as seen in figure \ref{jetgeo}, are open boundaries. The bottom boundary at $z_0 = 0.5$ pc is an open boundary for $r \geq r_b$ and was set to the jet injection for $r \leq r_b$, where $r_b = 0.05$ pc. The $r = 0$ boundary is the symmetry axis. We typically stopped the simulation before the jet reached the boundaries $z_{out}$ and $r_{out}$ to prevent energy from leaving the domain because we need to quantify energy deposition.

Fig. \ref{jetgeo} gives the geometry of the jet generated with a flaring-like scenario. In the initial conditions, the jet is considered to already have propagated for sometime and to have reached $z_1=1$ pc. The initial radius of the jet at inlet is fixed at $r_b=0.05$ pc and considers the jet to be coming out of a central source located at 0.5 parsec below the grid starting at $z_0$. Then according to the flaring-like scenario, introducing a fixed angle $\theta$, we can obtain the position of a virtual source at $z_v$, which for the angles of 5 and 10 degrees is somewhere in between $z=0$ and $z=z_0$,  as seen in Fig. \ref{jetgeo}. This was used to initiate the entire jet region within the computational domain, by considering the flux coming from this source and adopting a constant mass flux along the section of the cone. This means that we initialized density and velocity in this part of the domain with the following equations:

\begin{center}
\begin{equation}
\rho(r,z,t=0)=\rho_{\mathrm{jet}}\frac{r_b^2}{(r_b+(z-z_0)\tan(\theta))^2},
\label{}
\end{equation} 
\end{center}

\begin{center}
\begin{equation}
v_\mathrm{z}(r,z,t=0)=v~\cos(\arctan(\frac{r}{z-z_\mathrm{v}})),
\label{}
\end{equation} 
\end{center}

\begin{center}
\begin{equation}
v_\mathrm{r}(r,z,t=0)=v~\sin(\arctan(\frac{r}{z-z_\mathrm{v}})).
\label{}
\end{equation} 
\end{center}

\section{Results -parametric study}

\subsection{Dynamics - general description}
\label{general}

\begin{figure}
\begin{center}
  \includegraphics[width=.4\textwidth]{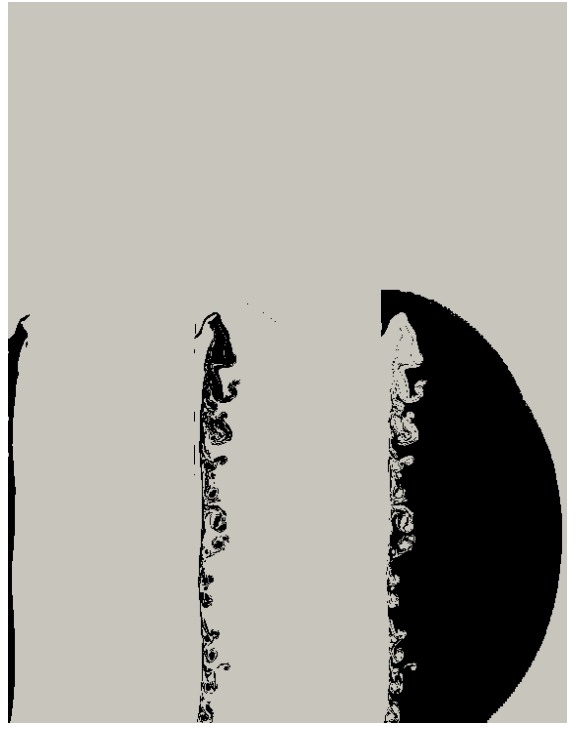}
	\caption{Example of our method to instantaneously identify jet regions for a jet opening angle of 10\textdegree~ at time t=20 (case E). We show the masks locating the jet (left) mixing region (center) and the shocked ISM (right). We identify the jet beam by $\gamma \geq 2$, finding the mixing region where $|(\nabla \times \vec{v})_{\phi}| > 1$ without the jet beam, and the cocoon where $\tau \neq \tau_{t=0}$ with beam and mixing region excluded.}
	\label{mask}
\end{center}
\end{figure}

As previously pointed out, we aim to study the transfer of energy between the jet and the ISM. To do this, we need to trace and quantify the significance and role of the different regions of the domain that characterize the jet beam, the instability-dominated region, the shocked ISM (SISM), and the unperturbed ISM.
The main feature of the region where the beam of the jet is located is its high value for the Lorentz factor. We empirically defined the threshold value to be $\gamma \geq 2$. With this criterion, the beam is well-captured without pollution from the cocoon region, which still has relativistic flows where $2 > \gamma > 1 $. The instability dominated region was located with the curl of the velocity field: because we did not include jet rotation and adopted jet axisymmetry, only the toroidal ($\phi$) component of the curl is non vanishing. We can instantaneously capture this mixing region where instabilities develop by locating where the magnitude of the curl component is higher than a threshold value of 1. The shocked region is identified by quantifying where the instantaneous energy density $\tau$ is not equal to its initial value. An example of these criteria applied to the simulation of case E at time 20 of the simulation can be seen in Fig. \ref{mask}. Because the regions are not suppose to overlap, we substracted the jet from the mixing region, and both jet and mixing region from the shocked ISM.

\begin{figure}
\begin{center}
	\includegraphics[width=.4\textwidth]{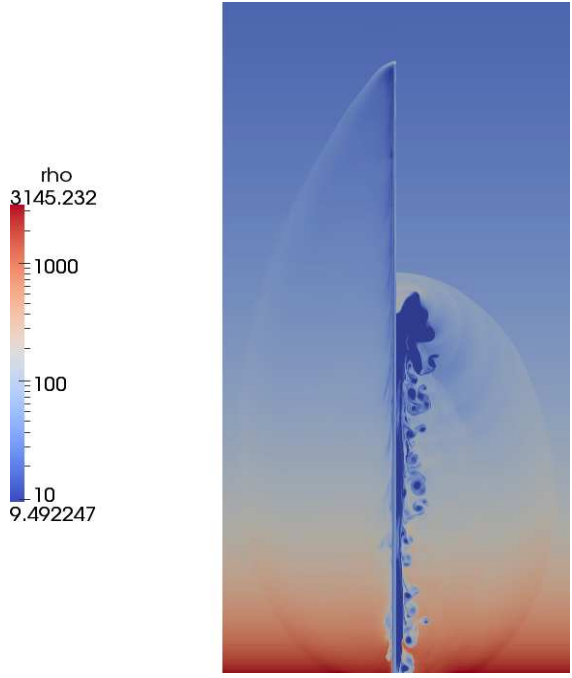}
	\caption{Comparison of the density distribution at time t=20 for an opening angle of 0\textdegree~ (left - case C) and 10\textdegree~ (right - Case E). A wider opening angle increases the width of the jet beam that stays collimated. The maximal reach of the jet beam is inversely proportional to the opening angle. The radial reach of the shock surrounding the jet remains of the same order.}
	\label{comparison0-10}
\end{center}
\end{figure}

\begin{figure}
\begin{center}
	\includegraphics[width=.4\textwidth]{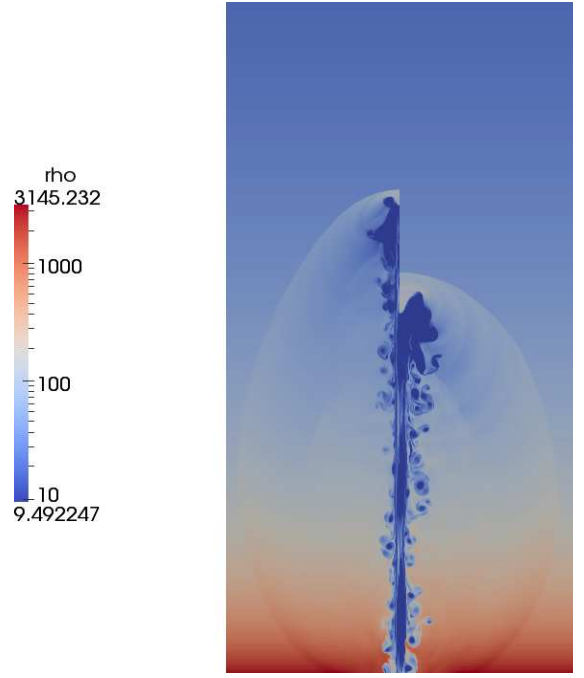}
	\caption{Comparison of the proper density distribution at time t=20 for an opening angle of 5\textdegree~ (left - case D) and 10\textdegree~ (right - Case E). Conclusions from visual investigation agree qualitatively with fig. \ref{comparison0-10}.}
	\label{comparison5-10}
\end{center}
\end{figure}

\begin{figure}
\begin{center}
	\includegraphics[width=.4\textwidth]{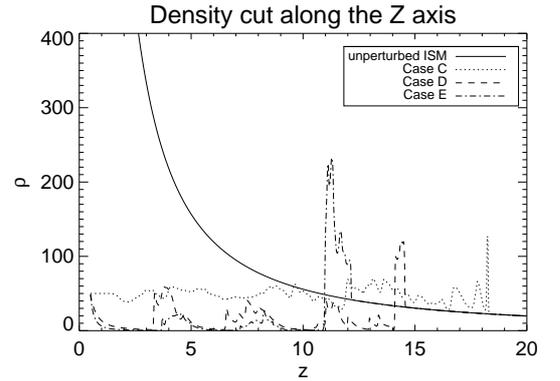}
	\caption{Density profile along the symmetry axis of the jet for cases C, D and E and for the unperturbed atmosphere. All cases show an increased density at the head of the jet beam. The maximal $z$ reach is inversely proportional to the opening angle. Case C (cylindrical case) keeps a near constant density. Cases D and E show a series of nodes with increased density. The position of the nodes depends on the opening angle: the wider the opening angle, the longer the internodal distance. The jet density in case C becomes higher than the ISM density from about $z\approx 12$, whereas for the other two cases, it remains below the ISM density profile throughout the domain.}
	\label{rhoprofile}
\end{center}
\end{figure}

Figs. \ref{comparison0-10} and \ref{comparison5-10} show instantaneous comparisons between cases C, D and E, with a 0\textdegree, 5\textdegree, and 10\textdegree opening angle with a King-type density profile, respectively. Common features can be described: the beam of the jet first shows an unperturbed region in which the matter propagated without interaction. It is then re-collimated and a shock forms that finally reaches the axis. Figure \ref{rhoprofile} shows the density variation along the symmetry axis at the same time $t=20$ for the three cases C, D and E. For the finite opening angle cases D and E, this plot shows that the position of this re-collimation shock (at about $z\approx 3$) slightly changes for varying opening angle. We discuss in section \ref{intern_structure} how these shocks, which are repeated along the symmetry direction, can provide a scenario for X-ray blobs.

\begin{figure}
\begin{center}
	\includegraphics[width=.4\textwidth]{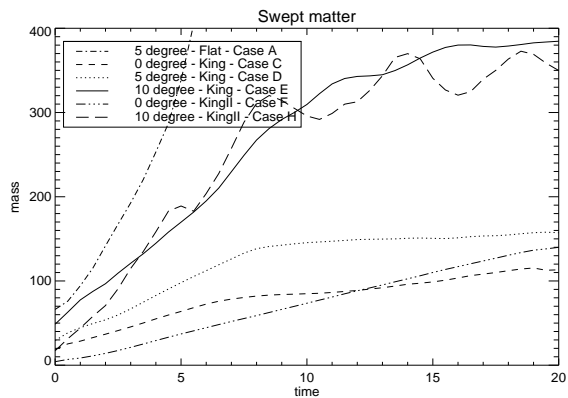}
	\caption{Matter swept up by the jet beam for different cases. For the same density profile (cases C, D and E, or cases, F and H) wider jets sweep away more matter. We can see that the density profile clearly matters. }
	\label{sweptedmatter}
\end{center}
\end{figure}

\begin{figure}
\begin{center}
	\includegraphics[width=.4\textwidth]{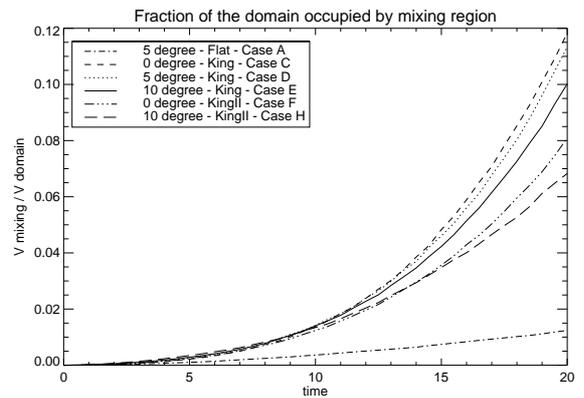}
	\caption{Volume fraction of the mixing region up to a time t=20 for different cases. The volume depends both on the opening angle and the properties of the ISM density profile. The volume increases for a smaller opening angle although the radial reach of this region diminishes. The maximal $z$ reach of the jet is therefore dominant over the radial expansion of the mixing region.}
	\label{volumeinstab}
\end{center}
\end{figure}

\begin{figure}
\begin{center}
	\includegraphics[width=.4\textwidth]{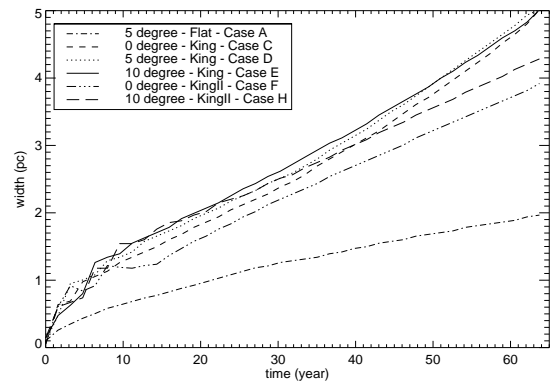}
	\caption{Radial expansion of the shocked region over time for different cases. Despite initial differences, the radial width of the jet-influenced region does not vary much for a varying opening angle.}
	\label{shockISM}
\end{center}
\end{figure}

\begin{figure}
\begin{center}
	\includegraphics[width=.4\textwidth]{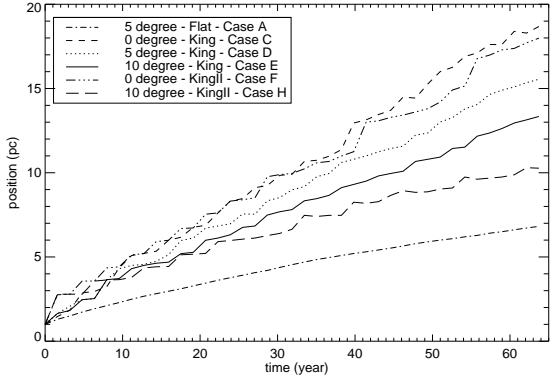}
	\caption{Position of the bow shock marking the head of the jet over time for different cases. The density ratio and the opening angle are both relevant for the propagation of this bow shock. The wider the opening angle, the more it decelerates. The higher the density ratio $\eta$, the slower the shock front deceleration.}
	\label{hpos}
\end{center}
\end{figure}

The head of the jet beam sweeps up matter and the wider the opening angle of the jet, the wider this head is and the more matter is pushed away. This qualitative statement can be quantified using the masks introduced earlier: we can estimate the mass swept up by quantifying 
\begin{equation} 
M_{\mathrm{swept}} = \iint_{\mathrm{beam}(t)} \rho_{\mathrm{ISM}}(r,z,t=0) \,2\pi r\,dr\,dz,
\end{equation}
which is shown in Fig. \ref{sweptedmatter}. A bow shock surrounds the whole structure and propagates into the ISM, transferring energy to an expanding region. By using the filters previously discussed, we performed a series of diagnostics: by integrating over the different regions, we can quantify the effect of the variation of the opening angle on, e.g. the fraction of the domain volume occupied by the mixing region, which we show in Fig. \ref{volumeinstab}. We can also instantaneously locate the overall position of the frontal bow shock as seen in the right panel of Fig.~\ref{mask}. Therefore we can quantify the maximal radial and axial expansion of this region over time as seen in Figs.~\ref{shockISM} and~\ref{hpos}. At the dynamically evolving contact interface between matter from the jet and the swept-up ISM, instabilities develop. These are of two kinds: Rayleigh-Taylor instabilities forming at the accelerated and decelerated contact interface of the jet as it propagates into the ISM, which then are advected along the side of the jet. Additionally, Kelvin-Helmholtz instabilities form along the jet beam flanks due to the velocity gradient between the moving matter of the jet that propagates axially and the slower matter from the interaction region with a more complex velocity field. The formation of both those instabilities can be followed with the filters discussed previously. We quantify how much of the energy transfer takes place when these instabilities mix matter from the jet and from the ISM. By looking at a plot of the curl of the velocity field, we can highlight those instabilities. As expected, at the interface between the matter of the jet with a high Lorentz factor and the more complex and slower matter from the interaction region, we can see a strong shearing of the fluid with high negative value of the curl of the velocity field. We can differentiate likewise the unperturbed part of the beam, and the part of the beam where recollimating shock waves are found. The front of these shock waves, as clearly identified from a pressure plot (not shown), coincide with a locally strong shearing. The vorticity mapping makes us able to track the instabilities and their evolution in time. By integrating the instantaneous energy density over the different regions as defined by the filters, we can obtain an estimation of energy transfer to the different regions.

\begin{figure}
\begin{center}
	\includegraphics[width=.4\textwidth]{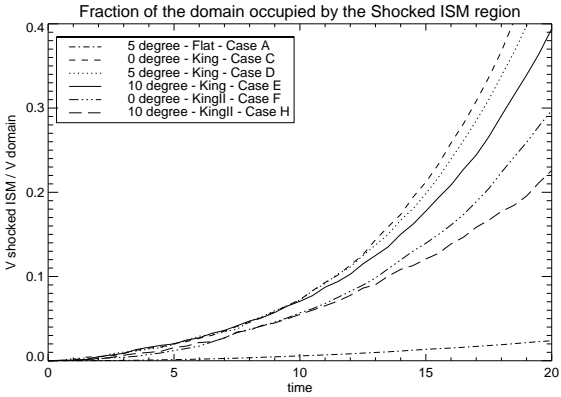}
	\caption{Fraction of the domain volume occupied by the shocked ISM region up to a time t=20 for different cases. This volume fraction of the shocked region both depends on the opening angle and on the density profile of the ISM. We know from Fig. \ref{shockISM} that all simulations have comparable radial reach, therefore we expect the volume of the cases to be ordered as in Fig. \ref{hpos}. This is the case within the same series of simulations (cases C, D and E, and cases F and H).}
	\label{volumeshockISM}
\end{center}
\end{figure}

\subsection{Dynamics - overall propagation distances}
\label{propagation}

As mentioned before, Fig.~\ref{hpos} shows the axial offset position of the bow shock front over time for an opening angle of the jet of 0\textdegree~ (case C), 5\textdegree~ (case D) and 10\textdegree~ (case E) with a King-type density profile as defined in section \ref{IC}. It shows the same for the pair of 0\textdegree~ (case F) and 10\textdegree~ (case H) with a King-II-type density profile. Finally, it quantifies the same position for an opening angle of 5\textdegree~ with a flat density profile (case A). All data sets start similarly as the head of the jet propagates at early time with the velocity obtained from the expression for pressure-matched jet propagation in 1D \citep{M.Marti1997}. Case A highlights how the density structure of the ISM affects the propagation. This uniform density case decelerates the head of the jet more as the influence of the ISM augments, in contrast to cases C, D and E, where the ISM density decreases. The inferred velocity of the head of the jet, quantified as the slope of the curves shown in Fig.~\ref{hpos}, has a low Lorentz factor around 1, and the velocity within the beam of the jet itself has a Lorentz factor of 10 or more (due to a compression of the flow, which can increase the Lorentz factor above its initial value). Different effects influence the acceleration or deceleration of the head: the principal one is the continuous ram pressure force exerted by the matter from the jet. As the jet beam propagates, the ram pressure force at the Mach disk is proportional to its surface area and depends on velocity. A significant factor is the accumulated mass of matter swept up by the jet beam. This was quantified earlier in Fig. \ref{sweptedmatter}. We discussed before that one of the main mechanisms of transfer of energy between the jet and the mixing region is dependent on the conversion from kinetic to thermal energy at the Mach disk, which is then mixed with the surroundings of the jet. Therefore, monitoring the velocity of the Mach disk, and the closely related propagation speed of the frontal bow shock as inferrable from Fig.~\ref{hpos}, gives us an indication on the efficiency of this energy transfer: one would expect that the kinetic energy from the jet would be transferred into thermal energy for the cocoon.

The upper limit of propagation is a ballistic behavior of no deceleration of the jet. The closest to this is the fully cylindrical jet of case C (0\textdegree, King), consistent with the least swept-up matter shown in Fig.~\ref{sweptedmatter}. The ram pressure felt by the head of the jet as it propagates decreases as the density of the ISM decreases in the axial direction. For cases D and E (5\textdegree and 10\textdegree King), we see a clear deceleration of the velocity of the head of the jet, due to a wider opening angle of the jet. Ultimately, we expect the mass of material piled up at the head of the jet to be high enough to stop its propagation, and this could happen sooner for wider opening angles and lead to shorter lengths of propagation (but this is on time-length scales beyond our simulations here). The test case scenario A with a flat density profile shows a strong deceleration as the jet encounters a constant high ISM density. As for the previous cases, the total mass of matter pushed aside by the jet increases, decelerating the head of the jet even more. This can be seen in Fig. \ref{hpos} where case A (5\textdegree, uniform density profile) shows the lowest curve. By comparison with case D (still 5\textdegree but King density profile) it becomes evident that the density profile of the ISM is a governing effect. All these results are consistent with Fig. \ref{sweptedmatter}: a wider opening angle increased the mass of matter pushed away and therefore implies a higher value of the ram pressure. For flat density profiles, the density does not decrease axially and we reach even higher values for the swept-up matter and therefore a much more significant deceleration of the head of the jet. From Fig. \ref{hpos} we can expect the same relation between the opening angle for the jet and the increase of energy transfer between the jet and the ISM. We also infer that a wider opening angle for the jet translates into a reduced reach of the jet into the ISM. This calls for longer simulations to study the longer term, far distance behavior.

Cases F and H (0\textdegree and 10\textdegree, King II density profile) are given in all figures as the limit cases of the second series of simulations entering a varying density atmosphere. The intermediate opening angle case G (5\textdegree, King II) is not quantified in the figures, but we verified that it is always intermediate between cases F and H. The conclusions obtained from the study of the second serie and their comparison between one another are similar to the ones obtained from the related series of cases C, D and E (0\textdegree, 5\textdegree, 10\textdegree, King). Case B (5\textdegree, uniform density profile, $\eta=10^{-4}$) is found to follow all trends that were identified with case A (5\textdegree, uniform density profile, $\eta=10^{-2}$), only shifted in intensity due to the difference of the density ratio $\eta$.

In summary: A wider opening angle results in shorter axial propagation, while the density of the medium in which the jet propagated plays the expected role: a higher density of the external medium results in a higher deceleration.

\subsection{Dynamics study - volumes and radial expansion}

As the jet propagates into the ISM, a bow shock front forms around it and propagates. This front stays always fairly close to the end of the jet beam, but radially it expands much faster than the jet which is re-collimated.
As this bow shock propagates into the ISM, it transmits energy to the local matter. This is the main way to transmit energy to far radial distances. 
Fig. \ref{shockISM} shows time evolution of the maximal radial distance reached by the SISM bounded by the bow shock, for several representative cases.

For propagation into a King atmosphere, we find that this maximal radial expansion of the SISM stays on the same order irrespective of the initial jet opening angle. While the jet beam on input is only 0.05pc for all cases, the bow shock's sideway influence region increases to 5 parsec within the simulation time.
For an opening angle of 5\textdegree, but with a flat density profile instead of a King-type atmosphere (case A), the radial expansion is slower and stays within 2 pc in the same timespan. We can easily explain this behavior by considering that this shock is propagating sideways with a speed influenced by the local sound speed. As said in section \ref{IC}, the initial pressure in the ISM is constant, whereas for case A, density is also constant, but density decreases with axial distance for other cases. Therefore the local sound speed (which is a factor of $\frac{p}{\rho}$) in the ISM increases in the axial direction. This in combination makes the sideways expansion more dependent on the ISM circumstances than on internal beam properties. The latter, as explained previously, does affect the beam axial propagation behavior, as shown in Fig. \ref{hpos}, where smaller opening angles meant more deceleration. Associated with this difference in reach in the axial direction for the jet, Fig. \ref{volumeshockISM} shows a higher volume fraction for the shocked region for smaller opening angles for the same density profile (compare in particular cases C, D and E, i.e. 0\textdegree, 5\textdegree and 10\textdegree, and King-type). This figure quantifies the ratio
\begin{equation}
\frac{V_{\mathrm{SISM}}}{V_{\mathrm{domain}}}=\frac{\iint_{{\mathrm{SISM}}(t)} 2\pi\,r\,dr\,dz}{\iint_{{\mathrm{domain}}} 2\pi\,r\,dr\,dz} \,.
\end{equation}
In reality, the pressure in the ISM is not constant as we move away from the source. Therefore to link those statements to observations, we would need to study the evolution of the ratio $\frac{p}{\rho}$ of the surroundings of AGN jets, which is still a problem for future study.

From Figs. \ref{comparison0-10} and \ref{comparison5-10}, one would expect the volume of the mixing region for cases C, D and E to be more significant for wider opening angle. But from Fig. \ref{volumeinstab}, which quantifies the instability dominated volume fraction, we can see that the radial expansion of the instability-influenced region is compensated for by the lesser axial propagation of the jet. By comparison, case A, which encounters a much higher external density, and therefore a much higher ram pressure, is confined to a smaller volume. For the jet beam region itself (as identified at each time with the mask as in the left panel of Fig. \ref{mask}), the main factor for the variation of its volume (figure not shown) is the radial expansion set by the initial opening angle: for an opening angle of 0\textdegree~ (case C), the jet does not expand radially and therefore its volume increases linearly with its axial expansion. For a wider opening angle of 5\textdegree~ and even more for 10\textdegree, the jet beam has a varying radial expansion due to recollimation by internal shocks and the varying ram pressure at its head.

In conclusion: The radial propagation of the frontal bow shock is dominated by the external medium parameters. This shock is the main means of energy transfer to large radial distances.

\subsection{Temporal evolution of the energy content}

\begin{figure}
\begin{center}
	\includegraphics[width=.4\textwidth]{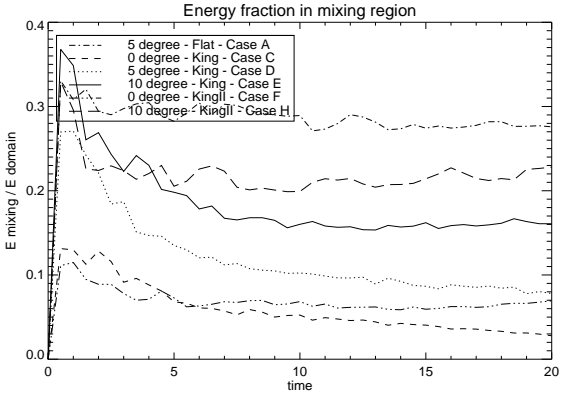}
	\caption{Fraction of the total domain energy excess present in the mixing region, which is dominated by instabilities, for the different cases.}
	\label{energyinstab}
\end{center}
\end{figure}

\begin{figure}
\begin{center}
	\includegraphics[width=.4\textwidth]{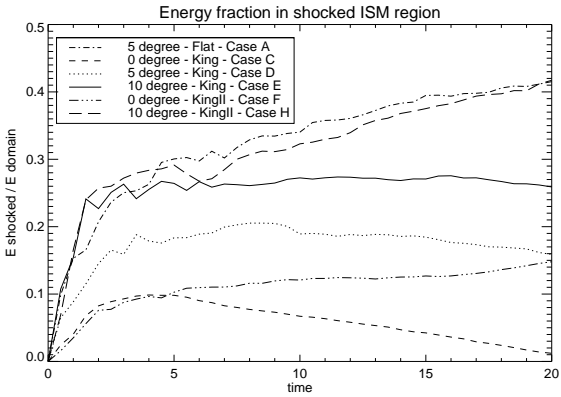}
	\caption{Excess energy ratio in the shocked ISM region for different cases.}
	\label{energypertub}
\end{center}
\end{figure}

\begin{figure}
\begin{center}
	\includegraphics[width=.4\textwidth]{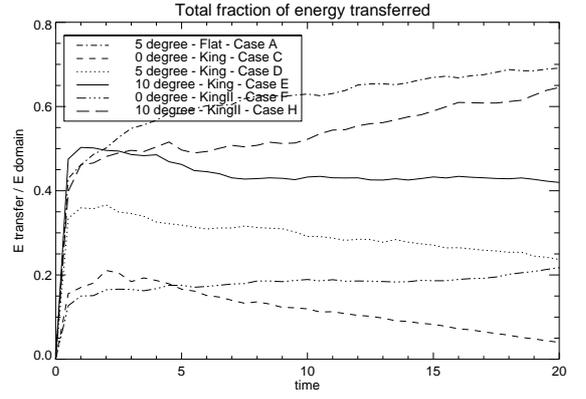}
	\caption{Total excess energy ratio in the mixing region and shocked ISM for different cases. By comparison with Figs.~\ref{energyinstab} and~\ref{energypertub} we conclude that the transfer is dominated by heating from the shocks in the shocked ISM region.}
	\label{energytransfer}
\end{center}
\end{figure}

We now discuss the evolution of the energy content by using various masks as described earlier in the text, to differentiate the contributions of different spatial regions (see section \ref{general}). We will do so by quantifying the fraction of excess energy present in each part of the domain as
\begin{equation}
\frac{E_{\mathrm{mask}}}{E_{\mathrm{domain}}}=\frac{\iint_{\mathrm{mask}(t)} \tau_{\mathrm{excess}}(r,z,t) \,2\pi r\,dr\,dz}{\iint_{\mathrm{domain}} \tau_{\mathrm{excess}}(r,z,t) \,2\pi r\,dr\,dz},
\end{equation}
where as excess energy we define the energy in each point of the domain, minus the contribution from the ISM at the same point for $t=0$:
\begin{equation}
\tau_{\mathrm{excess}}(r,z,t)=\tau(r,z,t)-\tau_{\mathrm{ISM}}(r,z,t=0)\, .
\end{equation}
For various choices of masks, we give the temporal evolution of the energy fraction present in the different parts of the domain, namely the mixing region (Fig.~\ref{energyinstab}), the shocked ISM (Fig.~\ref{energypertub}) and their combination (Fig.~\ref{energytransfer}). Note that the denominator is equivalent to the time integration of the jet influx after subtracting a small loss term $E_{\mathrm{loss}}$ that leaves through the open part of the bottom boundary:
\begin{eqnarray}
E_{\mathrm{domain}} &=&\int L_{\mathrm{jet}}\,dt -E_{\mathrm{loss}}\, \nonumber \\ 
&=& \iint_{\mathrm{domain}} \tau_{\mathrm{excess}}(r,z,t) \,2\pi r\,dr\,dz \, .
\end{eqnarray}
It is in particular non-zero at $t=0$ due to the finite region of the domain where we initialize the jet properties ($z\in[z_0,z_1]$ and $r$ within the specific conical segment). For the masks mentioned (mixing region and SISM), the numerator starts at a zero value.

\begin{figure}
\begin{center}
	\includegraphics[width=.4\textwidth]{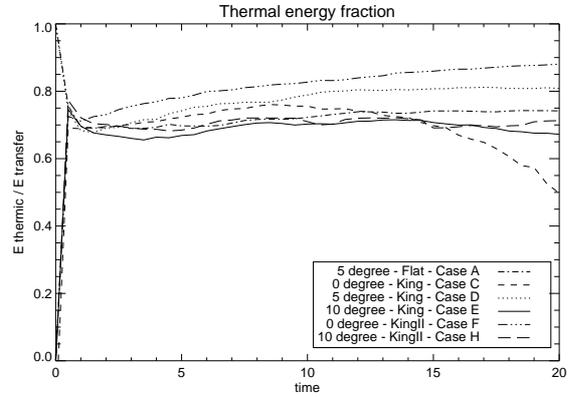}
	\caption{Thermal term of the total energy transfer.}
	\label{thermicfraction}
\end{center}
\end{figure}

For Figs. \ref{energyinstab} - \ref{thermicfraction}, the initial behavior up to time 1 in code units is mostly due to the way we initiate the simulation with a sharp-edged conical jet segment within the domain. More relevant for our discussion is the trend of these curves after this time.

In section \ref{propagation} we postulated that the energy transfer in the mixing region is dominated by the energy that is transferred through the end of the jet beam (Mach disk location). Therefore, a relation should exist between the total mass of swept up matter by the beam (as quantified earlier in Fig.~\ref{sweptedmatter}) and the energy transferred to the mixing region. 
Indeed, as we can confirm from Fig.~\ref{energyinstab},  within the same series of simulations (cases C, D and E: 0\textdegree, 5\textdegree and 10\textdegree, King or F and H: 0\textdegree, 10\textdegree, King II), by keeping the density profile fixed, the wider the opening angle, the higher the mass of swept-up matter and the excess energy transfer in the mixing region.
In all cases, we observe a decrease of the energy transfer rate to the mixing zone up to an asymptotic regime where the transfer becomes time-independent, e.g. cases E and H. Note that there is a clear trend for higher energy transfer for larger opening angles, minimizing the role of the ISM stratification.

Fig.~\ref{energypertub} shows the same quantification as before, with the mask now identifying the SISM regions. Similarly as before, wider opening angles result in higher energy transfer. In contrast to the mixing region case, the ISM stratification plays an important role in the energy transferred to the SIMS region. Indeed, Fig.~\ref{energypertub} shows a clear difference between cases C, D and E, and cases F and H. We point out that the energy transfer in the shocked ISM region is not dominated by the volumetric evolution of this region as seen in Fig.~\ref{volumeshockISM}. We can conclude that the dominant parameter for energy transfer is here the density profile of the ISM.

Combining both regions, Fig.~\ref{energytransfer} then quantifies how much energy is transferred to the jet beam surroundings as a whole. Contrary to our initial expectation that most of the energy transfer should occur at the Mach disk and mixing region alone, Fig.~\ref{energytransfer} shows that the total energy transfer is dominated by the transfer in the SISM region. We observe, on average, a ratio of 3 to 1 between the energy fraction present in the SIMS region, and its mixing region counterpart. Moreover, we identify the obtained percentage of the transferred energy as an estimate of the time-average energy transfer rate for the duration of our simulations,

\begin{equation}
\frac{\int L_{\mathrm{transfer}}(t) dt}{t} = \frac{E_{\mathrm{transfer}}(t)}{E_{\mathrm{domain}}(t)} \times L_{\mathrm{jet}} \,.
\end{equation}

In case D (5 degree opening angle with a King density profile) we observe a transfer rate of $2.5\times 10^{45}$ erg ${\mathrm{s}}^{-1}$ (i.e. at $t=20$, 25 \% of the total kinetic energy flux  $E_{\mathrm{domain}}$ is injected into the surrounding medium). We can quantify the ratio over thermal and kinetic energy to the total energy even better by noting that the local energy density minus rest mass, $\tau$, can be approximately expressed by its thermal and kinetic contribution as
\begin{center}
\begin{equation}
\tau_{\mathrm{thermal}} = p (\gamma^2\frac{\Gamma_{\mathrm{eff}}}{\Gamma_{\mathrm{eff}}-1}-1),
\end{equation}
\end{center}
\begin{center}
\begin{equation}
\tau_{\mathrm{kinetic}} = \gamma (\gamma-1) \rho c^2,
\end{equation}
\end{center}
where 
\begin{equation}
\Gamma_{\mathrm{eff}} =\Gamma - \frac{\Gamma-1}{2}(1 - \frac{m_p^2}{e^2})
\label{Geffequation}
\end{equation}
is the local varying effective polytropic index of the gas. In Fig.~\ref{thermicfraction} we use these individual contributions in the integral expressions, subtracting their initial ISM values.
We then find that about 80 percent of the energy transferred to the ISM is composed of the thermal term, which again confirms the dominant contribution of the bow-shock-related heating.

In conclusion: it is possible with our code to track where the energy is deposited and in which form, kinetic or thermal. For early stages of the propagation, the energy transfer is dominated by the heating by shocks which happens in the SISM region. We expect the transfer by mixing of the jet material and the surroundings to be dominant only at later times. It is also possible to quantify an average transfer rate of energy from the jet to its surroundings.

\section{Dynamical details in conical relativistic HD jets}

\subsection{Internal structure of the jet}
\label{intern_structure}

In Fig.~\ref{rhoprofile}, we compared the density profile along the symmetry axis for several cases. By overplotting the ISM density profile, it is possible to observe the evolution of the density ratio $\eta = \rho_{\mathrm{jet}} / \rho_{\mathrm{ISM}}$.
We recall that this ratio is set to 0.018 at the bottom boundary for several cases shown. For an opening angle of zero, the jet is cylindrical from the start of the domain and its density stays constant on average, whereas the density of the ISM decreases. Small fluctuations due to different internal shocks can be observed. For this case C (0\textdegree, King), after t=10 the density of the jet becomes higher than that from the ISM at a distance of 11 parsec from the central engine.

\begin{figure}
\begin{center}
	\includegraphics[width=.4\textwidth]{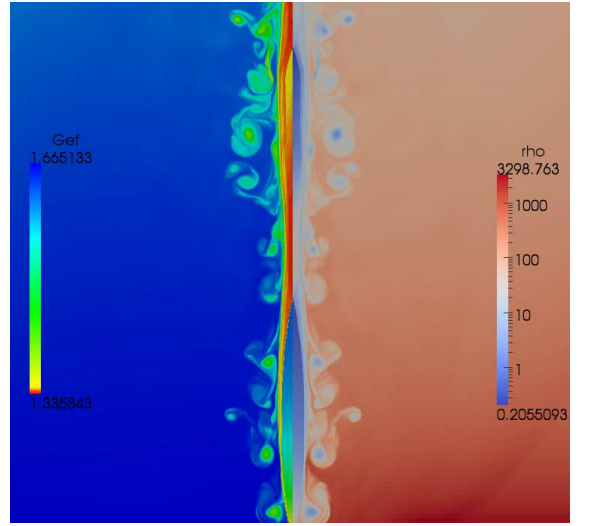}
	\caption{Details of two nodes along the axis. Internal structure of the jet beam appears with a finite opening angle: a static shock, in the lab frame, forms along the path of the jet, which can be the seat of Fermi-type-II acceleration of particles. The size of the image is $7\times 8$ pc with the two nodes at 3.5 and 7 pc along the symmetry axis.}
	\label{node}
\end{center}
\end{figure}

When finite opening angles are present, a clear recollimation shock forms.
Up to this recollimation shock, for an opening angle of 5\textdegree~ and 10\textdegree~ the jet density decreases as a factor of $\frac{1}{z^2 \tan( \theta)^2}$ with $z$ the distance to the source and $\theta$ the fixed opening angle of the jet. 
As $z$ increases, the recollimation shock wave propagates into the beam and compresses its density. 
The position at which this shock first reaches the symmetry axis evolves in time, until it finds an equilibrium recollimation position at 3 parsec for an opening angle of 5\textdegree, and 4 parsec for an opening angle of 10\textdegree. From time t=10 this quasi-equilibrium position is reached for both cases D and E, as in Fig.~\ref{rhoprofile} (similar trends emerge for the series F, G and H). A two-dimensional view of the near-stationary nodal structures is shown in Fig.~\ref{node} for case E at time $t=20$. This panel shows the density and effective polytropic index near the injection zone. 
Even if this does not dominate the dynamics, magnetic fields should be present in reality, enabling processes such as Fermi-II acceleration of particles in the nodes along the jet. This would result in an increase of the radiation processes at those positions, which is a scenario to explain the X-ray blobs found in observations of AGN jets.

\begin{figure}
\begin{center}
	\includegraphics[width=.3\textwidth]{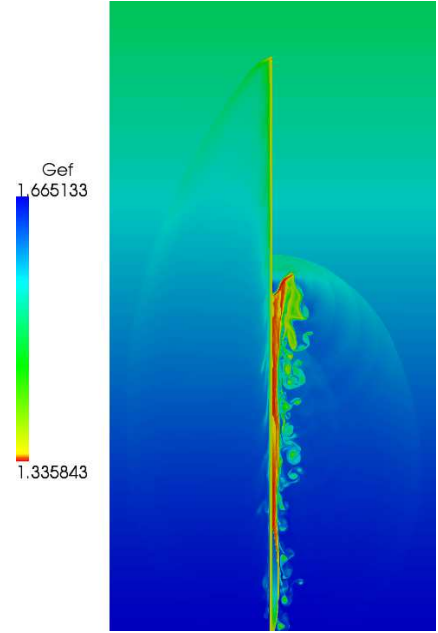}
	\caption{Comparison of the effective polytropic index for cases C (left) and E (right) at time 20 of the simulation. This figure reveals more clearly a series of secondary shocks that propagate in the shocked ISM region and also reheat it.}
	\label{Geff0-10}
\end{center}
\end{figure}

\begin{figure}
\begin{center}
	\includegraphics[width=.3\textwidth]{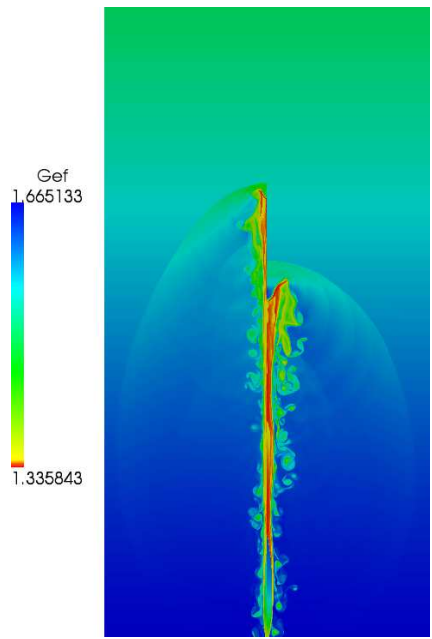}
	\caption{Comparison of the effective polytropic index for cases D (left) and E (right) at time 20 of the simulation.}
	\label{Geff5-10}
\end{center}
\end{figure}

We have a Synge-like closure relation, defined such that the effective polytropic index can vary between $5/3$, the classical polytropic index for a proton gas, to $4/3$, its relativistic value. We can recover the local value of this effective index and plot it following the equation \ref{Geffequation}.
This allows us insight into the thermodynamic condition of the gas. Figs. \ref{Geff0-10} and \ref{Geff5-10} show this effective polytropic index for cases C, D and E. We can see for both cases with a finite opening angle that the gas expanded and becomes cooler. This is due to adiabatic cooling in a freely expanding gas. The re-collimation shock reheats the gas and moves its polytropic index to the relativistic value of $4/3$. A zoom-in of this region was shown in Fig. \ref{node} for case E. The main jet beam remains at a relativistic temperature. For an opening angle of 0\textdegree~, the jet matter does not experience the adiabatic cooling phase and stays uniformly thermodynamically relativistic. We note that the core of the instability-related vortices also coincide with a mildly relativistic value of the polytropic index. In the variation of the effective polytropic index, we can also see how several secondary shocks traverse the SISM cocoon region, adding to the overall reheating.

In conclusion: using of an opening angle for the injection of the jet modifies its internal structure and gives rise to recollimation nodes along its path.

\subsection{Dynamic formation of a layered structure}

\begin{figure}
\begin{center}
	\includegraphics[width=.4\textwidth]{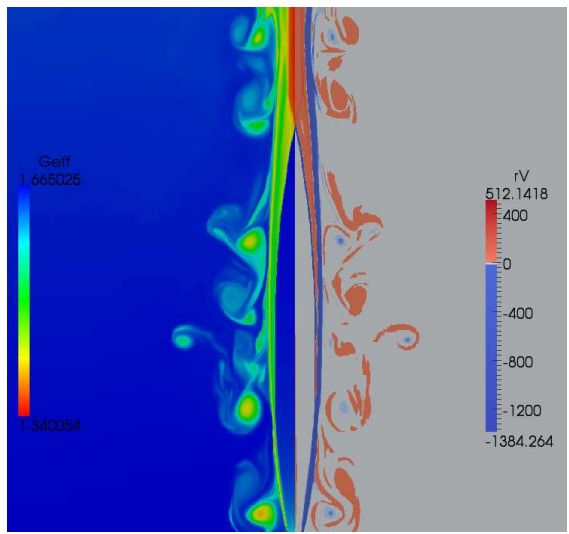}
	\caption{Presence of a dynamically formed layer surrounding the beam of the jet. This is another zoom on the inlet (as in Fig.~\ref{node}) for case E, now showing the (scaled) vorticity at the right.}
	\label{lowdensity}
\end{center}
\end{figure}

\begin{figure*}
\begin{center}
	\centering
	 \includegraphics[width=.4\textwidth]{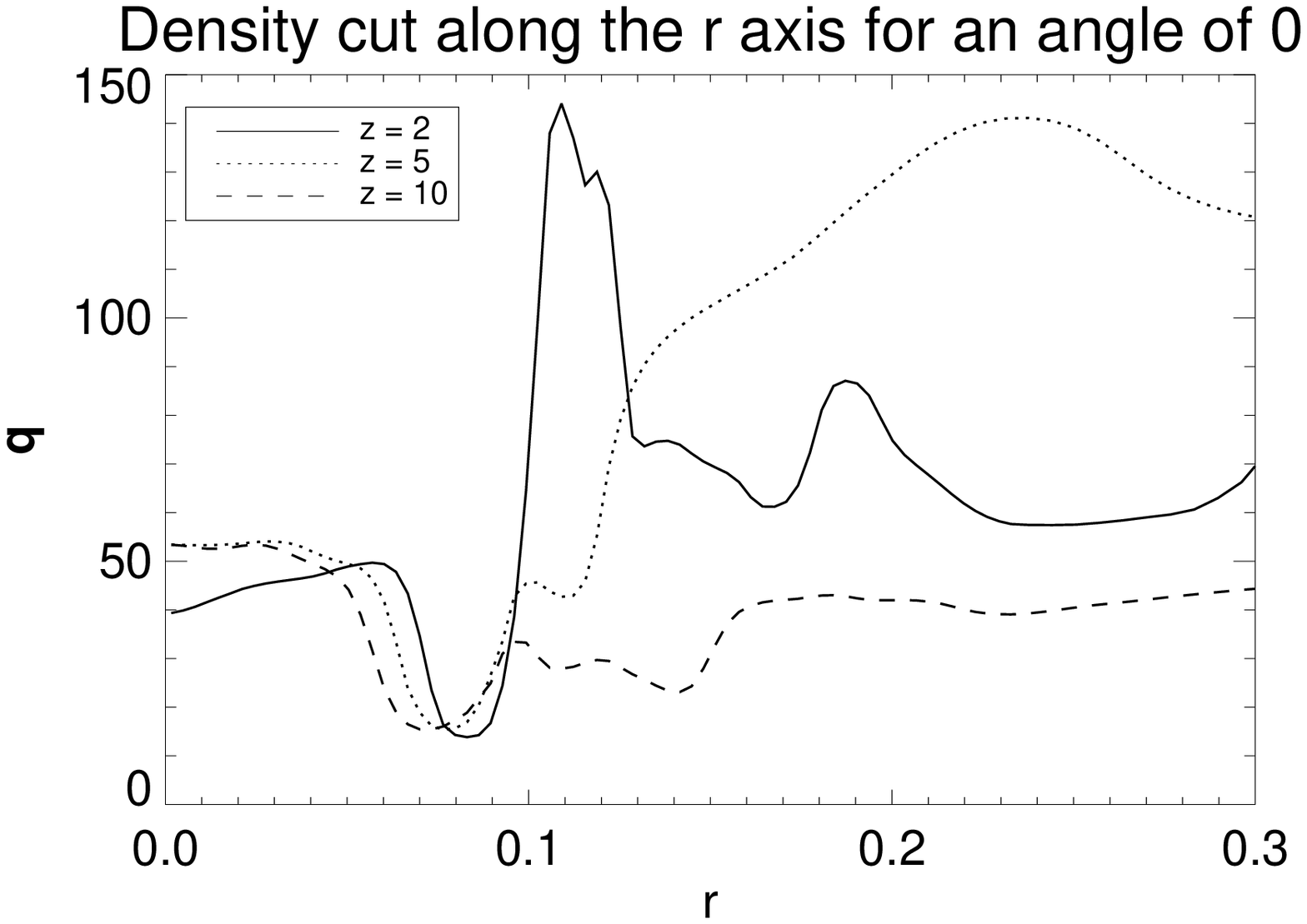}
	 \includegraphics[width=.4\textwidth]{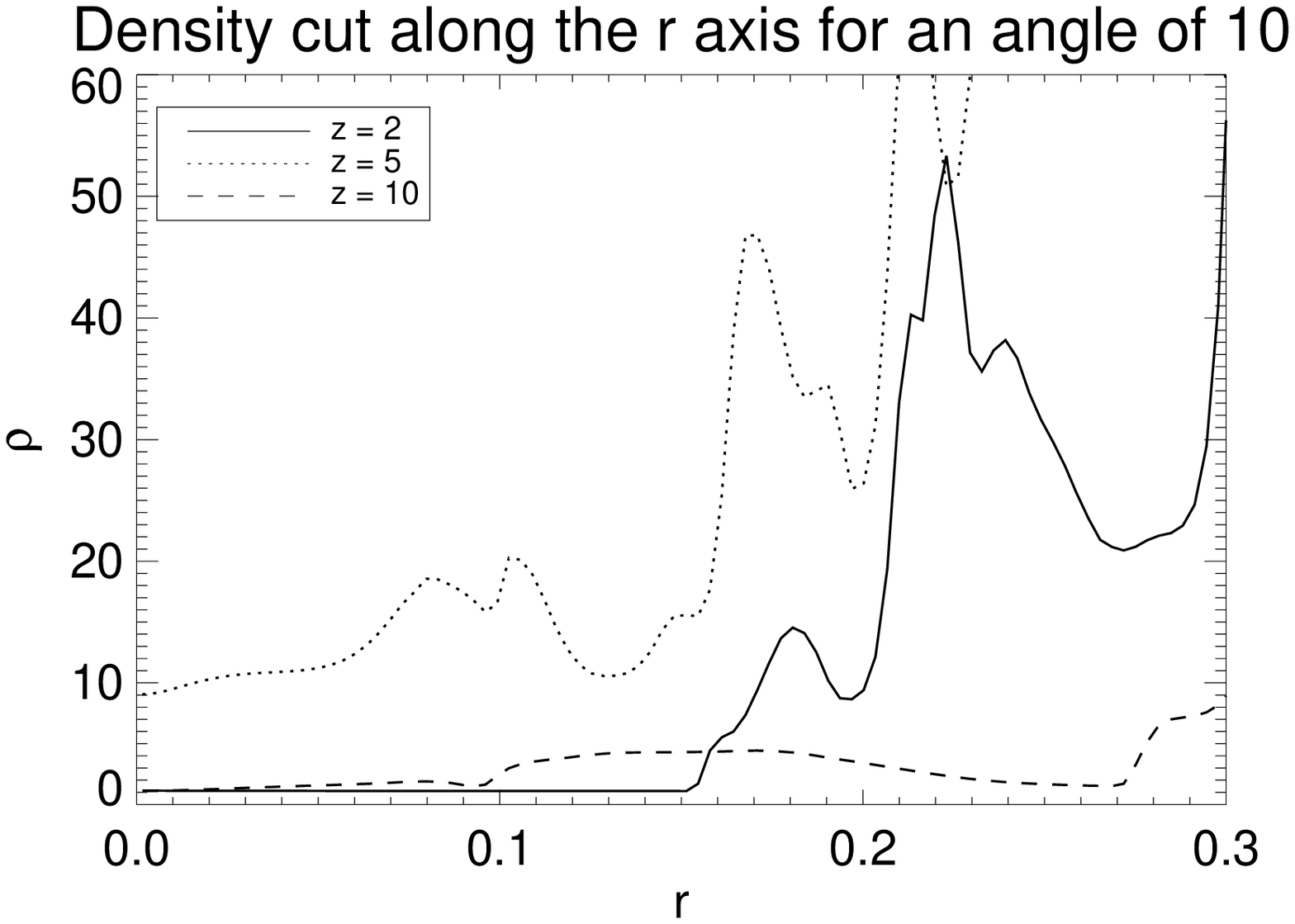}
	 \includegraphics[width=.4\textwidth]{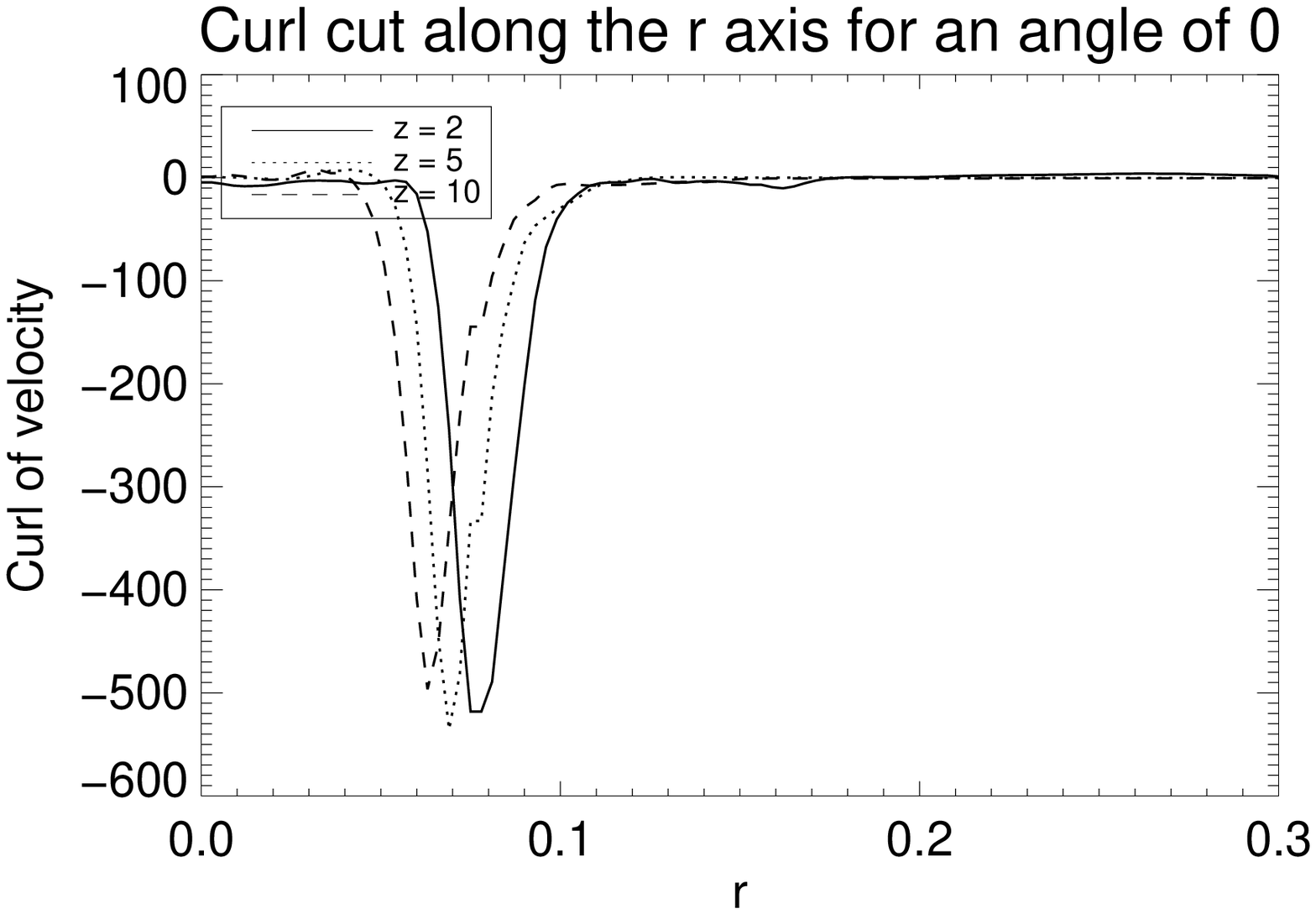}
	 \includegraphics[width=.4\textwidth]{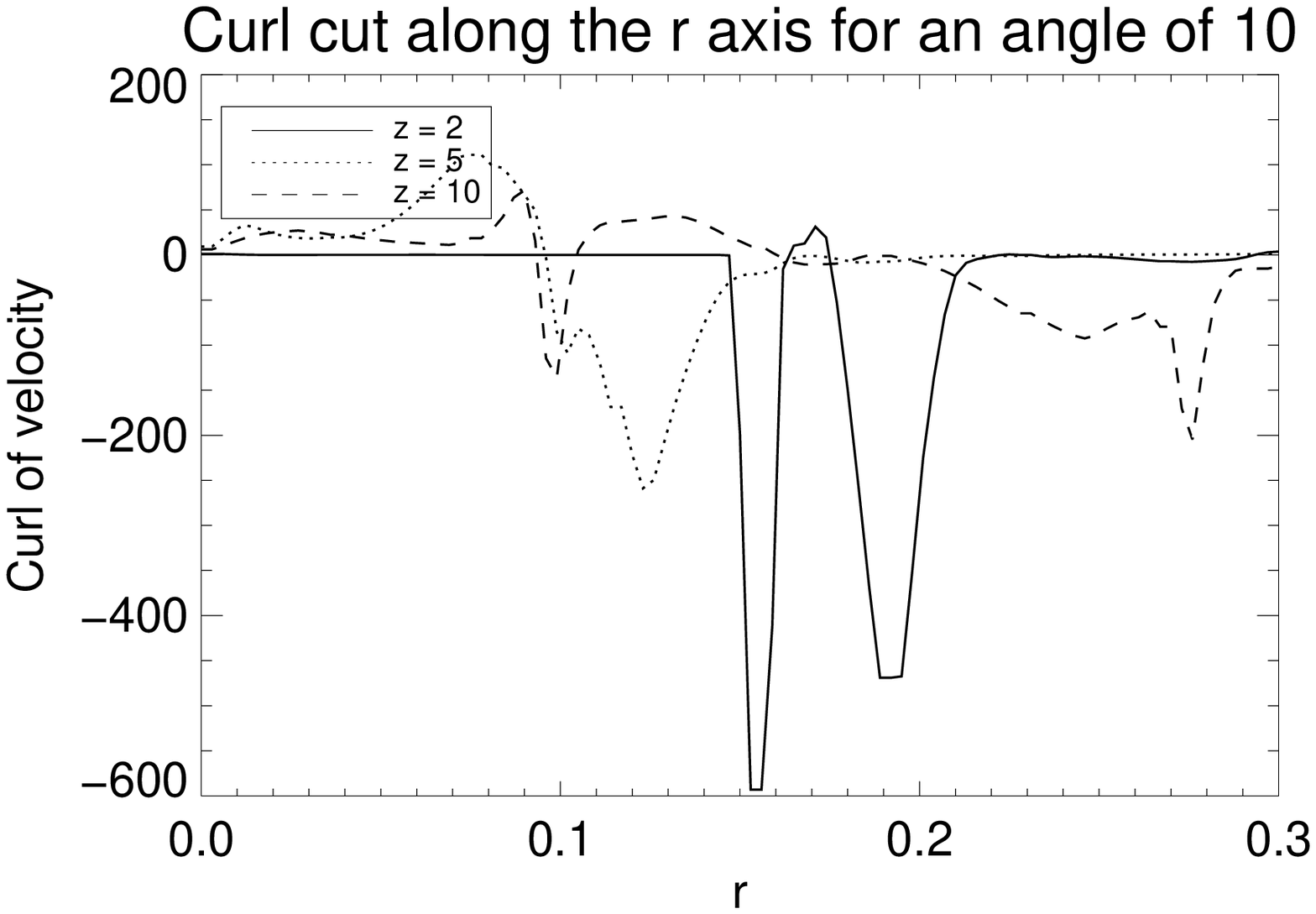}
	 \includegraphics[width=.4\textwidth]{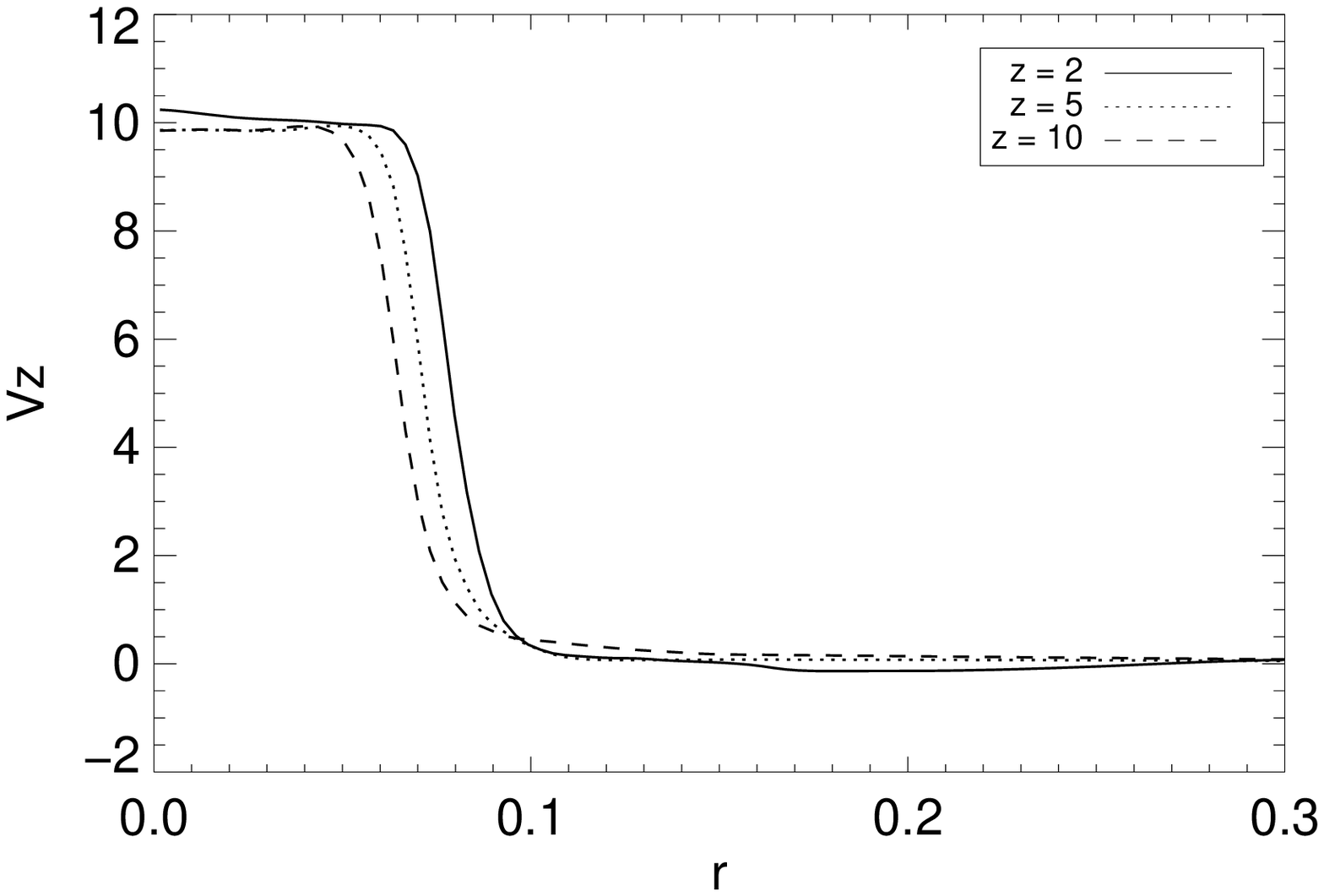}
	 \includegraphics[width=.4\textwidth]{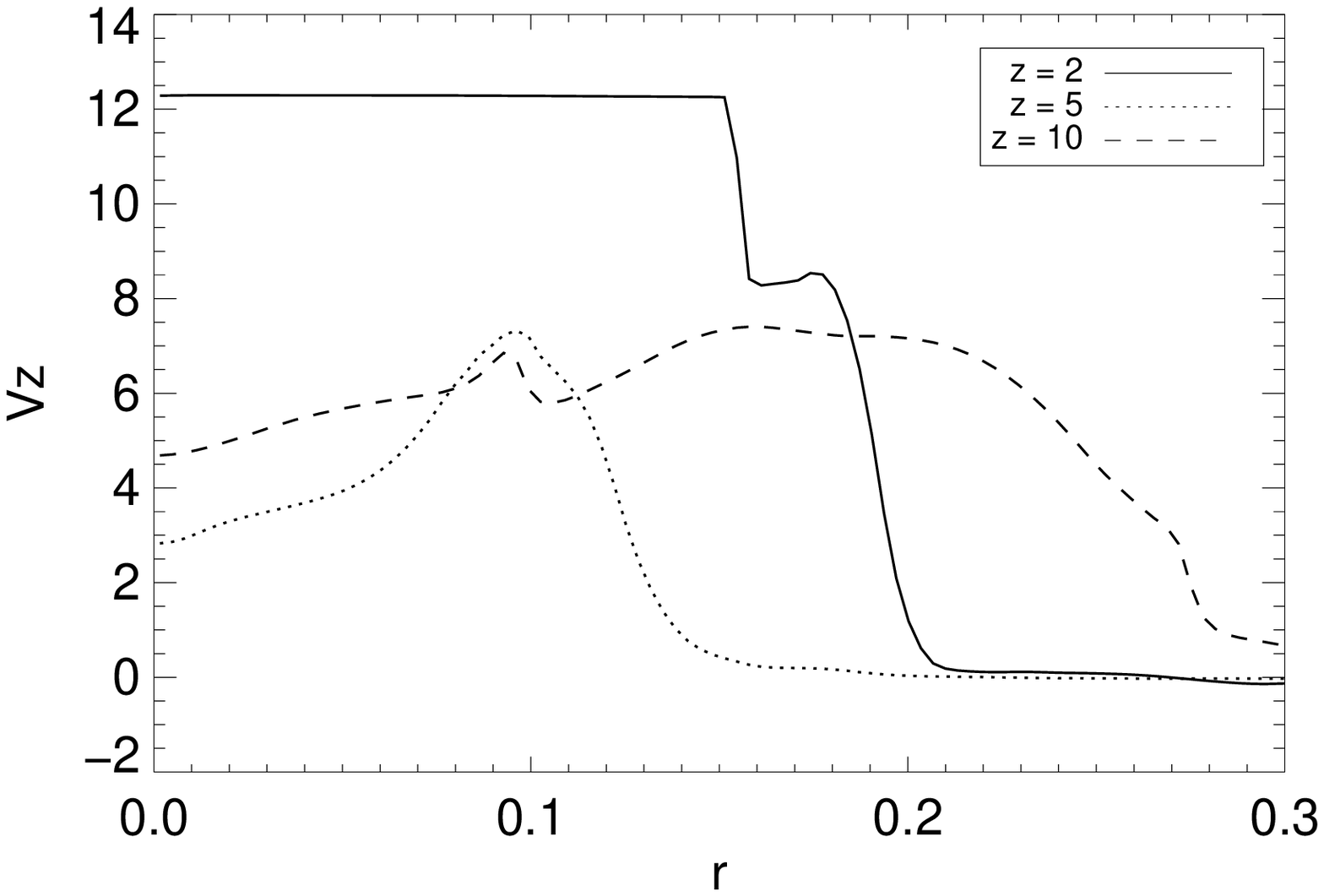}
	\caption{Radial cuts through the beam at distances z=2, 5 and 10 for an opening angle of 0\textdegree~ (left - case C) and 10\textdegree~ (right - case E), both with a King atmosphere density profile. Top - density; center - curl of the velocities; bottom - axial velocity. The cuts reveal the position of the jet beam and of a layer between the jet beam and the shocked ISM. For the cylindrical case (left) the layer is near constant in the symmetry axis direction and has a width comparable to the radius of the beam. The opening angle case (right) shows a more complicated structure of this dynamically formed layer, with two peaks in the curl of the velocity.}
	\label{cutbeam}
\end{center}
\end{figure*}

We also point out another detail of the finite opening angle cases by showing aanother zoomed view on the near inlet expansion region for case E (10\textdegree, King) in Fig. \ref{lowdensity}. We repeat on the left the local polytropic index, while the right panel now gives the local vorticity value. Especially the latter panel of Fig.~\ref{lowdensity} hints at the presence of a sharp radial structure in the jet beam, itself surrounded by the instabilities of the mixing region. Fig. \ref{cutbeam} shows a radial cut of the jet at fixed axial distances of 2, 5 and 10 parsec for cases C and E (0\textdegree and 10\textdegree, King). We can see at the interface between the jet and the shocked ISM that a layer forms with a width comparable with the beam width, with properties different from both media. For the cylindrical jet (opening angle of zero) this layer shows a lower density than the two surrounding regions with a clear local peak of the curl. For the 10 degrees opening angle we can see two such extrema in the local vorticity, present all along the jet from the root to the jet beam head, giving the beam a layered structure. These layered structures are sometimes invoked in studies of relativistic jets. \citet{ostrowski2001} used this layered jet to explain increased emission at the external border of the jet beam, whereas \citet{S.M.O'Neill2005} used it as a way to keep the jet in equilibrium with its surroundings. We recall that in our initial condition, the variation from the density of the jet to the density of the ISM is a step function. A sideways traveling rarefaction wave induces the shape we see eventually all along the entire jet beam boundary at later times. Case E is more complicated because the beam is surrounded by the growing instabilities and because of re-collimation, the edge of the beam is not radially static in space anymore. 

In conclusion: a radially layered jet structure is present also without an opening angle for the injection of the jet, but the latter does amplify this structure in a finite rarefaction zone.

\section{Conclusion}

We quantified the fraction of energy present in the different jet-affected regions for realistic FR-II-type AGN jets. It was possible to show that the main mechanism of energy transfer from the jet to the ISM is through heating via shocks in the shocked ISM region, on top of mixing-related energy exchange. The energy transfer we measured for the SISM region dominates the same measurement for the mixing region with a ratio of 3 to 1 on average. It is important to note that in reality, the mechanisms studied here coexist with magnetic energy transfers and radiation transfer. 

We discussed the role of finite opening angles on the dynamics of the jet, and in terms of efficiency of energy transfer. A wider opening angle increases the energy transfer but reduces the reach of the jet in its main direction of propagation and also reduces the portion of space that receives the energy. Our study followed relativistic Lorentz factor 10 jets for a lifetime of about 65 years and for a propagating distance up to 20 pc. We found that the amount of energy feedback from the jet to the ISM can vary from 10\% to 70 \% across all cases studied, where a total $10^{46}$ erg.s$^{-1}$ influx was assumed throughout. The jets were followed-up to asymptotic regimes where some of the energy transfer rates reached stationary values, which calls for larger scale, longer-time simulations to confirm these trends. This is because we know that astrophysical AGN jets propagate on scales of kpc to Mpc for extended lifetimes. These longer time and larger scale runs can be found in the works of \cite{perucho2011}. They found that for early times, the energy transfer from the jet to its surroundings is dominated by heating by shocks before mixing processes become dominant for later times. This matches our results where for early times of the jet we found a clear dominance of the SISM heating by shocks.\cite{perucho2011} adopted a resolution of 50 pc at 1 kpc from the source, assuming a cylindrical propagation of the jet making it complementary to our own simulations. While their study reveals the long-term, long-distance trends with merely two gridpoints in the jet inlet beam, our study quantifies the effect of finite opening angles for early-time, beam-resolved cases.

we also provided details on two dynamic phenomena due to finite opening angles. The first are the nodes corresponding to recollimation shocks forming along the jet path. These nodes can be seats of significant acceleration of particles, which in turn sets a scenario to explain the X-ray nodes in observations of AGN jets. This was originally studied in~\cite{komissarov1997}, with the results of relativistic hydro simulations translated in synthetic images. The opening angle of the jet plays an important role for the formation of these nodes, while the precise position and number of the nodes depends on the value of the opening angle.
The second phenomenon we observed was the dynamically formed layered structure of the jet. Starting with a sharp profile for the density that separates the jet and its surrounding, the beam becomes surrounded by a clear density-structured layer of precisely the form needed to explain spine-sheath scenarios that are usually included for jet stability arguments or emission-related inferences. This feature does not need a finite opening angle to appear, but a higher opening angle does change its properties, such as a wider radial width and structured velocity profile.

The next step of this work is to obtain synthetic X-ray maps of these simulations to compare them directly with observations. In this paper we intentionally focused on dynamical effects and energy transfer quantifications only. Moreover, we performed a parametric survey to identify clear trends due to opening angle variations and/or ISM density profiles. In future works, we intend to target specific well-diagnosed radio sources, and extend our study to the more computationally challenging 3D large-scale scenarios.

\begin{acknowledgements}
Thanks to Oliver Porth, Alkis Vlasis and Jan Deca for their support and help.
For part of the simulations, we used the infrastructure of the VSC -
Flemish Supercomputer Center, funded by the Hercules Foundation and the
Flemish Government - department EWI. We acknowledge financial support
from FWO-Vlaanderen, grant G.0238.12 and from project GOA/2009/009 (KU
Leuven).
\end{acknowledgements}

\bibliographystyle{aa}
\bibliography{article}

\end{document}